\documentclass[twocolumn,preprintnumbers,amsmath,amssymb]{revtex4}
\usepackage{graphicx}% Include figure files
\usepackage{dcolumn,comment}% Align table columns on decimal point
\usepackage{bm}% bold math

\newcommand{\be}{\begin{equation}}
\newcommand{\ee}{\end{equation}}
\newcommand{\bea}{\begin{eqnarray}}
\newcommand{\eea}{\end{eqnarray}}

\newcommand{\beq}{\begin{equation}}
\newcommand{\eeq}{\end{equation}}
\newcommand{\cD}{\mathcal{D}}
\newcommand{\cL}{\mathcal{L}}

\newcommand{\cN}{\mathcal{N}}

\newcommand{\cA}{\mathcal{A}}
\newcommand{\cH}{\mathcal{H}}

\newcommand{\cF}{\mathcal{F}}

\newcommand{\kcs}{k_{\rm cs}}
\newcommand{\bbZ}{\mathbb{Z}}

\newcommand{\cref}{{\bf [check ref]}}

\usepackage{wasysym}

\begin{document}
\preprint{MPP-2012-92}
\title{A Kaluza-Klein inspired action for chiral $p$-forms and their anomalies
}% Force line breaks with
\author{Federico Bonetti, 
        Thomas W.~Grimm, 
        Stefan Hohenegger}
\affiliation{  Max Planck Institute for Physics, \\
               F\"ohringer Ring 6, 80805 
               Munich, Germany 
\\
}
\begin{abstract}

The dynamics of chiral $p$-forms can be captured by a lower-dimensional parity-violating action 
motivated by a Kaluza-Klein reduction on a circle.
The massless modes are $(p-1)$-forms with standard kinetic terms and Chern-Simons couplings to the Kaluza-Klein vector of the background metric. 
The massive modes are $p$-forms charged under the Kaluza-Klein vector and admit parity-odd first-order kinetic terms. 
Gauge invariance is implemented by a St\"uckelberg-like mechanism using $(p-1)$-forms.
A Chern-Simons term for the Kaluza-Klein vector is generated at one loop by massive $p$-form modes. 
These findings are shown to be consistent with 
anomalies and supersymmetry for six-dimensional supergravity theories 
with chiral tensor multiplets.

\end{abstract}
%\pacs{...}% PACS, the Physics and Astronomy
\maketitle

%%%%%%%%%%%%%%%%%%%%%%%%%%%%%%%%%%%%%%%

\section{Introduction and Summary}

Chiral $p$-forms are $p$-forms with self-dual or anti-self-dual field strength.
They play an important role in string theory and M-theory. For instance, 
the massless spectrum of Type IIB superstring theory contains a 
chiral four-form, and the worldvolume theory of an M5-brane includes 
a chiral two-form (tensor). 
From a field-theoretic point of view, quantization of such fields 
is a non-trivial task, since it is notoriously hard to impose the duality
constraint at the level of the action \cite{Marcus:1982yu}. Different solutions to this problem 
have been proposed, based on breaking of manifest Lorentz invariance, introduction of auxiliary fields, or
a holographic approach \cite{Siegel:1983es}.  

In this work we consider chiral $p$-forms
in $D=2p+2$ dimensions with Lorentz signature. This  
requires $p$ to be even. A circle compactification 
leads us to a $(D-1)$-dimensional action
which can be used to study the 
dynamics of these $p$-forms. 
This approach is inspired by recent insights 
into string and M-theory effective actions. Firstly, 
six-dimensional (2,0) superconformal field theories for a 
stack of M5-branes have been conjectured to be equivalent to five-dimensional
super Yang-Mills theories \cite{Douglas:2010iu}.
Secondly, six-dimensional effective actions of  
F-theory compactifications with an arbitrary number of chiral tensors
have been derived by using the 
dual five-dimensional M-theory setups \cite{Bonetti:2011mw}.
In both frameworks excited Kaluza-Klein modes are essential 
for the correspondence between the six- and five-dimensional physics.

Our starting point is a $D$-dimensional pseudoaction, which
has to be supplemented by the self-duality constraint 
at the level of the equations of motion, in the spirit of
e.g.~\cite{Bergshoeff:2001pv}. One spatial direction is compactified on a circle,
and chiral $p$-forms are expanded onto a Kaluza-Klein tower 
of $(D-1)$-dimensional $p$- and $(p-1)$-forms. Both zero-modes and excited modes are retained,
and are subject to duality constraints coming from self-duality 
in $D$ dimensions.
These constraints can be implemented in a proper $(D-1)$-dimensional action,
which is
given explicitly in eq.~\eqref{final_total} below.
Zero-modes are described by massless $(p-1)$-forms, while
excited modes are described by $p$- and $(p-1)$-forms related
by a St\"uckelberg-like gauge symmetry. Accordingly, $p$-forms
can become massive by absorption of $(p-1)$-forms. The resulting
action \eqref{gauge_fixed_excited} generalizes the action studied in
\cite{Townsend:1983xs}. In particular, we couple the system to the
Kaluza-Klein vector. This proves to be crucial for our discussion.

Our classical action \eqref{final_total} 
can be quantized by standard methods. We claim that this provides
us with a window into the $D$-dimensional quantum theory. 
As a highly non-trivial check, 
we consider as an example $D=6$, and 
compute a particular one-loop
Chern-Simons coupling in the five-dimensional theory.
The result is compared against independent predictions
from six-dimensional (1,0) and (2,0) supergravities. In particular, 
in the former case we are able 
to extract information about gravitational anomalies
of the original six-dimensional theory
first computed in \cite{AlvarezGaume:1983ig}.

We expect our formalism to be also useful
in the study of other systems with duality constraints. For instance, it may be applied to
the democratic formulation of Type II supergravities \cite{Bergshoeff:2001pv} or
to four-dimensional Maxwell actions with manifest electric-magnetic duality, see e.g.~\cite{Schwarz:1993vs}.

%%%%%%%%%%%%%%%%%%%%%%%%%%%%%%%%%%%%%%%

\section{Lower-dimensional action for chiral $p$-forms}

A free chiral $p$-form $\hat B$ in $D=2p+2$
dimensions  (with $p$ even) is subject to the self-duality condition
\begin{equation} \label{minimal_self}
 \hat * \, \hat \cH = c \, \hat \cH \ ,
\end{equation}
where $c=\pm 1$ and $\hat \cH = d\hat B$. 
This constraint  is first-order, and is not easily derived from an action. However, differentiation of \eqref{minimal_self}
gives a second-order equation
which is readily obtained from the pseudoaction
\begin{equation} \label{minimal_pseudo}
 \hat S =  \int - \tfrac 14 \hat \cH \wedge \hat * \, \hat \cH \ .
\end{equation}
The prefactor is chosen to have canonical normalization
in the following discussion.
Note that the pseudoaction formalism can be also applied to setups including several $p$-forms and their couplings
to other fields. 

Let us now put the pseudoaction \eqref{minimal_pseudo} on a circle, by means of the Kaluza-Klein 
ansatz for the metric,
\begin{equation} \label{ansatz_metric}
 d\hat s^2(x,y) = ds^2(x) + r^2(x) [dy+A^0(x)]^2 \ .
\end{equation}
In this relation, $x$ are the non-compact $D-1$ coordinates, $y \sim y + 2\pi$ is the coordinate
along the circle, $r$ is the compactification radius, and $A^0$ is the Kaluza-Klein vector, with field
strength $F^0 = dA^0$.

We expand the $D$-dimensional $p$-form $\hat B$ in Kaluza-Klein modes according to
\begin{equation} \label{ansatz_Bfield}
 \hat B = \sum_{n \in \bbZ} e^{iny} \big[  B_{n} + A_{n} \wedge (dy+A^0) \big ]\ ,
\end{equation}
where $B_{n}, A_{n}$ are $(D-1)$-dimensional $p$-forms and $(p-1)$-forms, respectively,
and only depend
on the non-compact coordinates $x$. Our formalism requires $p>0$,
and hence is not applicable to chiral scalars in two dimensions. Note that Kaluza-Klein modes are subject to
a reality condition, e.g.~$\bar B_{n} \equiv (B_{n})^* =  B_{-n}$.

Dimensional reduction of the higher-dimensional field strength $\hat\cH$
is conveniently described in terms of the lower-dimensional field strengths
\begin{equation} \label{GcF_expr}
 \cH_{n}  = \cD B_{n} - A_{n} \wedge F^0  \ , \quad
 \cF_{n}  = \cD A_{n} + i n B_{n} \ ,
\end{equation}
where we have introduced the covariant exterior derivative
$\cD = d - i n A^0$ acting on the $n$th mode.
Note in particular the St\"uckelberg-like coupling
in the second equation, which ensures invariance under 
\begin{equation} \label{shift_symmetry}
 \delta B_{n} = \cD \Lambda_{n} \ , \quad
\delta A_{n} = - i n \Lambda_{n} \ .
\end{equation}

A straightforward computation shows that the pseudoaction \eqref{minimal_pseudo}
is reduced to the sum $\sum_n \tilde S_{n}$, where
\begin{equation} \label{excited_action}
 \tilde S_{n} = \int -\tfrac 14  r\, \bar\cH_{n} \wedge *\, \cH_{n} -\tfrac 14 r^{-1} \bar\cF_{n} \wedge *\, \cF_{n}  \ .
\end{equation}
 Finally, the self-duality constraint \eqref{minimal_self}
yields a constraint for each Kaluza-Klein level, $ r * \cH_{n} = c\, \cF_{n}$.
In the following, we implement these constraints at the level of the lower-dimensional action. 
To this end, zero-modes and excited modes are treated differently.

For the sake of simplicity, we will henceforth drop the Kaluza-Klein subscript on zero-modes, $B\equiv B_0$, $A \equiv A_0$.
As we can see from \eqref{shift_symmetry}, the shift symmetry of the theory acts trivially on the zero-mode
$A$. Because of the self-duality
constraint, $B$ and $A$ thus furnish a redundant description of the same degrees of freedom,
and no gauge-fixing condition can eliminate this redundancy. Therefore, either $A$ or $B$ has to be
eliminated by hand from the action. 
In the following, we choose to remove $B$ and construct an action in terms of $A$ only.

To achieve this goal, we modify $\tilde S_{0}$ given in \eqref{excited_action} adding
\begin{equation} \label{zero_correction}
 \Delta \tilde S_{0} =  \int \tfrac 12 c \, \cH \wedge \cF + \tfrac12 c\;  A^0 \wedge \cF \wedge \cF \ .
\end{equation}
This term is a total derivative as a functional of $A,B,A^0$, and is such that the sum
$\tilde S_{0} + \Delta \tilde S_{0}$ can be written as a functional of $A,\cH, A^0$. Moreover, \eqref{zero_correction}
is engineered to get the duality constraint  $r * \cH = c\, \cF$ upon variation with respect to $\cH$, which 
appears only algebraically. We are thus able to integrate out $\cH$ to get 
a proper $(D-1)$-dimensional action depending on $A,A^0$ only. It reads 
\begin{equation} \label{final_zero_action}
 S_{0} =  \int -\tfrac 12 r^{-1} \cF \wedge * \cF  + \tfrac12 c\;  A^0 \wedge \cF \wedge \cF \ .
\end{equation}
Note that \eqref{GcF_expr} implies $\cF=dA$ for $n=0$.

Let us now turn to the discussion of the self-duality condition for the $n$th excited modes $B_{n}, A_{n}$.
For $n\neq 0$, the shift symmetry \eqref{shift_symmetry} acts non-trivially on $A_{n}$. As a result,
the redundancy of the formalism is simply a manifestation of gauge invariance.
Both $B_{n}$ and $A_{n}$ are thus allowed to enter the action in the gauge-invariant combination $\cF_n$
given in \eqref{GcF_expr}. 

The distinctive feature of the $n\neq 0$ case is the identity $ \cD \cF_{n} = i n \cH_{n}$,
which is immediately derived from \eqref{GcF_expr}. It  
allows us to modify $\tilde S_{n}$ in \eqref{excited_action} by adding
\begin{equation} \label{excited_correction}
 \Delta \tilde S_{n} =  \int
\tfrac 14 c\;  \bar\cH_{n} \wedge \cF_{n} + \tfrac{i}{4n} c\;  \bar\cF_{n} \wedge \cD \cF_{n}  + {\rm c.c.}  
\end{equation}
Indeed, this quantity is a total derivative as a functional of $A_{n}, B_{n}, A^0$.
However, the total action $\tilde S_{n} + \Delta \tilde S_{n}$ can be seen as a functional of 
$\cF_{n}, \cH_{n}, A^0$, in which $\cH_{n}$ enters only algebraically. As in the discussion
of the zero-modes, the duality constraint $r * \cH_{n} = c\, \cF_{n}$ is implemented through 
integrating out $\cH_{n}$. We are thus left with the proper action 
\begin{equation} \label{final_excited}
 S_{n} =  \int - \tfrac12 r^{-1} \bar\cF_{n} \wedge * \cF_{n} + \tfrac{i}{2n} c\;  \bar\cF_{n} \wedge \cD \cF_{n} \ ,
\end{equation}
where $A_{n}, B_{n}$ only appear through $\cF_{n}$.

We are now in a position to write down the total action in $D-1$ dimensions. 
It reads
\begin{multline} \label{final_total}
 S =  \int -\tfrac 12 r^{-1} \cF \wedge * \cF  + \tfrac12 c\;  A^0 \wedge \cF \wedge \cF \\
+  {\textstyle \sum\limits_{n=1}^\infty} \int  - r^{-1} \bar\cF_{n} \wedge * \cF_{n} + \tfrac{i}{n} c\;  \bar\cF_{n} \wedge \cD \cF_{n} \ .
\end{multline}
Note that we sum \eqref{final_excited} over positive $n$ only,
thanks to the reality conditions on $A_{n},B_{n}$.

It is worth pointing out that the physical degrees of freedom 
of excited modes can be described in terms of a massive $p$-form $B_{n}$ only.
In fact, the gauge symmetry \eqref{shift_symmetry} can be fixed imposing the condition $A_{n}=0$,
thus setting $\cF_n = in B_n$. As a result, the second line of \eqref{final_total} becomes
\begin{equation} \label{gauge_fixed_excited}
 {\textstyle \sum\limits_{n=1}^\infty} \int  - n^2 r^{-1} \bar B_{n} \wedge * B_{n} + i c n\;  \bar B_{n} \wedge \cD B_{n} \ .
\end{equation}
The classical mass is $m_n=(n^2 r^{-1})(cn)^{-1}= cn r^{-1}$.

Note that \eqref{gauge_fixed_excited} is invariant under local U(1) transformations
of the complex $p$-form $B_n$ gauged by $A^0$. In 
\cite{Townsend:1983xs} this gauging is absent, and therefore it is possible to
integrate out the real or imaginary part of $B_n$ consistently. The resulting action is the standard massive Proca action for $p$-forms
and has no explicitly parity-violating terms. By contrast, the gauging in \eqref{gauge_fixed_excited}
introduces parity-odd interactions that are
essential for our analysis.

The action \eqref{gauge_fixed_excited} is expected to be supersymmetrizable in many cases of interest, since
our findings are reminiscent of tensor hierarchies in supergravity.
For $\cN=2$ models in five dimensions, we refer the reader to e.g.~\cite{Bergshoeff:2004kh}. 

Note that \eqref{gauge_fixed_excited} has strong analogies with the Lagrangian
of Kaluza-Klein modes $\chi_{n}$ of a higher-dimensional spin-$1/2$ fermion on the circle,
\begin{equation} \label{excited_fermion}
 \cL_{n}^{\rm ferm} = - \bar\chi_{n} \gamma^\mu \cD_\mu \chi_{n} + m_n \bar\chi_{n} \chi_{n}  + \cL_{n}^{\rm supp} \ .
\end{equation}
First of all, $\cD_\mu = \partial_\mu - i nA^0_\mu$ contains minimal coupling to $A^0$ with charge $n$. Second of all, 
 the lower-dimensional mass parameter $m_n = c_{1/2} n r^{-1}$ depends on the higher-dimensional
chirality $c_{1/2}$. Finally, in $\cL_{n}^{\rm supp}$ couplings are collected which are suppressed by the mass scale $r^{-1}$. 
Couplings of this kind will play a key role in the quantum theory of massive forms studied in the next section.

%%%%%%%%%%%%%%%%%%%%%%%%%%%%%%%%%%%%%%%

\section{Quantization and one-loop test}

The action \eqref{final_total} 
can be quantized using standard methods.
More importantly we claim that this allows us to derive physical statements about 
the underlying higher-dimensional theory at the quantum level. 
In order to provide evidence for this, we show that a one-loop 
computation in the quantized theory \eqref{final_total} 
is consistent with known features of the underlying higher-dimensional 
theory. More precisely, we will compute 
the contributions of massive $p$-forms $B_{n}$ to the Chern-Simons couplings of the form 
$A^0 (F^0)^p$ in the $(D-1)$-dimensional effective theory. 
For concreteness we will focus on chiral two-forms in six dimensions
and analyze the coupling 
\beq \label{SCS000}
     S^{(5)}_{\rm cs} =  -\tfrac{1}{48} \int \kcs \, A^0 \wedge F^0 \wedge F^0 \ .
\eeq

We can provide a heuristic, diagrammatic argument to show that \eqref{SCS000} is sensitive to six-dimensional anomalies.
 Consider an anomalous four-graviton
one-loop amplitude in six dimensions, and choose the polarization tensors 
in the external legs in such a way to extract the component
$\langle \hat g_{yy} \hat g_{\mu y} \hat g_{\nu y} \hat g_{\rho y} \rangle$,
where $y$ is the compact coordinate. As can be seen from \eqref{ansatz_metric}, 
this four-point function in six-dimensions is related to $r \langle A^0_\mu A^0_\nu A^0_\rho \rangle$ in five dimensions,
once the metric component $\hat g_{yy}$ is replaced by its background value $r$.
In six dimensions, the anomalous part of the amplitude is generated by 
massless chiral fields running in the loop.
In five dimensions, we are thus led to compute the contribution to $\langle A^0_\mu A^0_\nu A^0_\rho \rangle$
 coming from all Kaluza-Klein modes of these chiral fields.

For supergravity theories our results are compatible with higher-dimensional 
gravitational anomalies \cite{AlvarezGaume:1983ig}, and vanishing conditions imposed by supersymmetry \cite{Awada:1985ep}.

In six dimensions gravitational anomalies are induced 
by chiral two-forms and chiral spinors.
We will consider first $\cN=(1,0)$ supergravities. 
The relevant multiplets are
the gravity multiplet containing one left-handed USp(2)-doublet of gravitinos 
and one self-dual two-form, $H$ hypermultiplets containing a right-handed spin-$1/2$ fermion,
$V$ vector multiplets containing a left-handed spin-$1/2$ fermion, 
and $T$ tensor multiplets containing an anti-self-dual two-form 
and a right-handed spin-$1/2$ fermion. 

Quantum consistency of the theory requires cancellation of the 
gravitational anomalies. In six dimensions this generically requires a 
Green-Schwarz term \cite{Green:1984sg} for the $T+1$ chiral two-forms 
$B^\alpha$ of the form 
\beq \label{GS-term}
   \tfrac{1}{2} \int a^\alpha \Omega_{\alpha \beta} B^\alpha \wedge \text{Tr} (R \wedge R) \ ,
\eeq
where $a^\alpha,\Omega_{\alpha \beta}$ are a constant vector and matrix, respectively,
and $R$ is the curvature two-form in six dimensions.  
In the six-dimensional $(1,0)$ theories the gravitational anomalies are cancelled if \cite{Sadov:1996zm}
\beq \label{gravitational_anomalies}
 H-V=273-29 T\ ,\qquad a^\alpha \Omega_{\alpha \beta} a^\beta = 9-T\ .  
\eeq

In a next step one performs a Kaluza-Klein reduction of this $(1,0)$ theory on
a circle to five dimensions, with details found in \cite{Bonetti:2011mw,BonettiGrimmHohenegger}. 
We consider the 
five-dimensional Chern-Simons coupling \eqref{SCS000}.
Using the Kaluza-Klein ans\"atze \eqref{ansatz_metric}, \eqref{ansatz_Bfield} a direct computation yields that the classical part of 
$\kcs$ is actually vanishing. It is, however, suggested in \cite{Bonetti:2011mw,BonettiGrimmHohenegger}  
that this term is induced by integrating out massive Kaluza-Klein modes.
In fact, using M-/F-theory duality one expects $\kcs= a^\alpha \Omega_{\alpha \beta} a^\beta$ 
\cite{Bonetti:2011mw}. 
By comparison with \eqref{gravitational_anomalies} one then infers that 
\beq \label{prediction10}
  \kcs = 9-T\ . 
\eeq
We aim to get this result from a five-dimensional one-loop computation. 
In fact, $\kcs$ can be read off from the 
parity violating part of the three-point amplitude $\langle A^0 A^0 A^0\rangle$
bilinear in the momenta. As discussed in \cite{BonettiGrimmHohenegger}, massive modes of chiral tensors, chiral spin-$1/2$ fermions, 
and chiral spin-$3/2$ fermions contribute. 
Here we report some details about the tensor contribution only and refer to
\cite{BonettiGrimmHohenegger} for a complete account of the full computation.

From the gauge-fixed action \eqref{gauge_fixed_excited}
 we determine the propagator $\langle B_n^{\phantom{m}\lambda\tau} {\bar B}_{m\, \mu\nu}\rangle$, which is proportional to 
\beq \label{propagator}
   \delta_{mn}\frac{ 2 \delta_{[\mu}^{[\lambda} \delta_{\nu]}^{\tau]} - 4 m_n^{-2} k_{[\mu} \delta_{\nu]}^{[\lambda} k^{\tau]} + m_n^{-1} \epsilon_{\mu \nu \rho}^{\phantom{\mu \nu \rho}\lambda \tau} k^\rho }{{k^2+m_n^2}}\ ,
\eeq
where $k^\mu$ is the momentum, and $m_n$ is the tensor mass introduced after \eqref{gauge_fixed_excited}.
From \eqref{gauge_fixed_excited} we can also read off a vertex proportional 
to $n \epsilon^{\mu \nu \rho \lambda \sigma}$, which will be referred to as electric 
vertex in the following.  

\begin{figure}[htbp]
\centering
\includegraphics[width=6 cm]{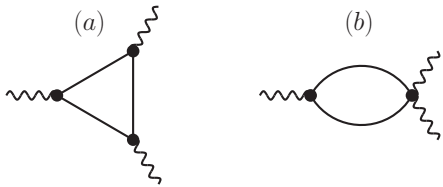}
\caption{Diagrams contributing to $\langle A^0 A^0 A^0 \rangle$.\label{fig:diagrams}}
\end{figure}

The diagram we compute first is the loop diagram (a) of figure \ref{fig:diagrams}
with three electric vertices. This diagram by itself is
divergent and requires regularization. 
Using the propagator \eqref{propagator} and the electric vertex,
naive power counting yields a cubic UV divergence in the momentum cutoff.  
Explicit computation of the parity violating term, however, shows that this 
term only diverges linearly \cite{BonettiGrimmHohenegger}. 
This linear divergence can be cancelled by 
adding the two counter-terms
\beq \label{counter}
    \bar B^{\mu}_{\phantom{\mu} \nu} B^{\nu \rho} F^0_{\rho \mu}\, \ ,\quad  \bar B^{\mu \nu} F^0_{\nu \rho} B^{\rho \lambda} F^0_{\lambda \mu}\ ,
\eeq
where the Kaluza-Klein level $n$ has been suppressed on all $B_{\mu \nu}$.  
These couplings introduce new vertices which modify the contribution from 
diagrams of type (a) and introduce new diagrams of type (b) depicted in 
figure \ref{fig:diagrams}. 
Note that one counterterm is not sufficient since each of them 
introduces a cubic divergence into the parity-violating part of the amplitude. Thus, two parameters
are needed to cancel all divergences.

In order to present the full result of the computation of
the parity-violating part of $\langle A^0 A^0 A^0 \rangle$, 
we introduce the notation $\cA^{X}_{n}$, where $X$ indicates the type of field running in the 
five-dimensional loop at the $n$th Kaluza-Klein level. The values of $\cA_{n}^X$
for tensors, spin-$1/2$ fermions, and spin-$3/2$ fermions are respectively \cite{BonettiGrimmHohenegger}
\beq \label{blocks}
  \cA^{B}_{n} = - 4 c n^3\, , \ \cA^{1/2}_{n} =  c_{1/2} n^3\, , \ \cA^{3/2}_{n} =  5 c_{3/2} n^3\, ,
\eeq 
where a common normalization has been fixed and the coefficients $c_X=\pm 1$ indicate the chirality.
Note that the mass scale $r^{-1}$ drops from the computation, so that 
the only dependence on the Kaluza-Klein level is contained in the common dimensionless
factor $n^3$. 

Summing all contributions \eqref{blocks} from chiral fields in the spectrum of (1,0) supergravity
 we get
\beq \label{total10}
\cA_{n}^{\rm (1,0)} = n^3 \left[ -4(1-T) + 2(V - H -T) + 10 \right]\ ,
\eeq
where fermionic contributions receive an extra factor of 2 since they carry
an USp(2) index. Precise matching between $\sum_{n} \cA^{(1,0)}_n$ and $\kcs$ requires a suitable 
regularization of the divergent sum $\sum n^3$. Independent of this normalization issue,
if the first relation in \eqref{gravitational_anomalies} is  
imposed in \eqref{total10}, $\cA_{n}^{\rm (1,0)}$ is proportional to $9-T$.

Note that the counterterms \eqref{counter} have mass dimension
greater than five and are suppressed by the compactification
mass scale $r^{-1}$. Similarly to get \eqref{total10}, divergences of the 
fermionic diagrams have been 
cancelled using $r^{-1}$-suppressed  counterterms. 
Thus this renormalization scheme is suitable for Kaluza-Klein reductions.

Finally, let us also briefly discuss the analog situation in 
a $(2,0)$ theory. In this case, the gravity multiplet comprises two gravitinos 
and five self-dual tensors, and each of the $T$ tensor multiplets includes 
one anti-self-dual tensor and one USp(4) right-handed spin-$1/2$ fermion. If
\eqref{blocks} are summed over this spectrum, one has
\begin{equation}
 \cA^{(2,0)}_{n} = n^3 \left[ -4(5-T) - 4 T + 20  \right] \equiv 0 \ ,
\end{equation}
which is consistent with the fact that the 
Chern-Simons coupling $A^0 F^0 F^0$ is
forbidden in any five-dimensional theory with 16 supercharges \cite{Awada:1985ep}.

%%%%%%%%%%%%%%%%%%%%%%%%%%%%%%%%%%%%%%%

%%%%%%%%%%%%%%%%%%%%%%%%%%%%%%%%%%%%%%%

{\noindent \bf Acknowledgments}

We would like thank Michael Douglas, Olaf Hohm, and Wati Taylor for useful discussions.

\end{document}